\documentclass[twocolumn,aps,prb,amsmath,amssymb,floatfix]{revtex4}
\usepackage[latin9]{inputenc}
\usepackage{amsmath}
\usepackage{amssymb}

\makeatletter
\@ifundefined{textcolor}{}
{%
 \definecolor{BLACK}{gray}{0}
 \definecolor{WHITE}{gray}{1}
 \definecolor{RED}{rgb}{1,0,0}
 \definecolor{GREEN}{rgb}{0,1,0}
 \definecolor{BLUE}{rgb}{0,0,1}
 \definecolor{CYAN}{cmyk}{1,0,0,0}
 \definecolor{MAGENTA}{cmyk}{0,1,0,0}
 \definecolor{YELLOW}{cmyk}{0,0,1,0}
}

\usepackage{tabularx}\usepackage{bm}\usepackage{euscript}\usepackage{color}\usepackage{amsfonts}\usepackage{exscale}\usepackage{amsbsy}\usepackage{subfigure}\usepackage{dcolumn}

\def\be{\begin{equation}}
\def\ee{\end{equation}}
\def\ba{\begin{eqnarray}}
\def\ea{\end{eqnarray}}

\def\C60{A$_x$C$_{60}$}

\makeatother

\begin{document}

\title{Kerr effect as evidence of gyrotropic order in the cuprates - revisited}

\author{Pavan Hosur$^{1}$}

\author{A. Kapitulnik$^{1}$}

\author{S. A. Kivelson$^{1}$}

\author{J. Orenstein$^{2,3}$}

\author{S. Raghu$^{1,4}$}

\author{W. Cho$^{1}$}

\author{A. Fried$^{1}$}

\affiliation{$^{1}$Department of Physics, Stanford University, Stanford, CA 94305}

\affiliation{$^{2}$Department of Physics, University of California, Berkeley,
CA 94720}

\affiliation{$^{3}$Materials Science Division, Lawrence Berkeley National Laboratory,
Berkeley, CA 94720}

\affiliation{$^{4}$SLAC National Accelerator Laboratory, Menlo Park, CA 94025}
\begin{abstract}
Recent analysis 
 has confirmed earlier
general arguments that the Kerr response vanishes in any time-reversal
invariant system which satisfies the Onsager relations. Thus, the
widely cited relation between natural optical activity (gyrotropy)
and the Kerr response, employed in Hosur \textit{et al},
Phys. Rev. B \textbf{87}, 115116 (2013), is incorrect.  However, there is increasingly clear experimental evidence
that, as argued in our paper, the onset of an observable Kerr signal
in the cuprates reflects 
point-group symmetry rather than time-reversal
symmetry breaking. 
\end{abstract}

\date{\today}

\maketitle
Measurements of the rotation of the 
polarization of 
normal-incidence light upon reflection (Kerr effect) provide an extremely useful probe of any  order which breaks time-reversal symmetry and all mirror  symmetries, {\it e.g.} ferromagnetic order with a component of the moment perpendicular to the surface.  
However,  
arguments have been put forward that a combination  of 
gyrotropy (handedness) and dissipation can lead to a Kerr response, even in the absence of  time-reversal symmetry breaking.
  In Hosur {\it et al.}\cite{Hosur2013} we adopted 
  these arguments and on this basis interpreted  the onset of the Kerr effect below a well-defined, doping-dependent onset temperature in cuprate superconductors in terms of spontaneous generation of handedness induced by charge ordering. As outlined below, we are now persuaded by a review of the literature and by our own explicit calculations, that Kerr rotation is forbidden  for a system subject to the combined constraints of linear response, thermal equilibrium, and time-reversal symmetry. We therefore wish to retract our suggestion that chiral charge density wave order, by itself, provides a basis for understanding the Kerr rotation observed in the cuprates.

Assuming linear response, the Kerr effect can be inferred from the dielectric 
tensor of the material.  It was a key element of the design of the Sagnac Interferometer employed\cite{spielman1990,spielman1992,kapitulniknew} in the experiments in question, that it should reject all ``non-reciprocal" effects.
Reciprocity relations are implicit in linear response theory, which were used by Halperin \cite{Halperin1992} to 
demonstrate that the Kerr effect requires time-reversal symmetry breaking beyond simple dissipation.  However, controversy existed before the Halperin paper\cite{Agranovich1972,Agranovich1973,Schlagheck1973,Schlagheck1975,Silverman1986,Silverman1990,Silverman1992} and continued to linger after it was published\cite{Bungay1993,Bungay1993a,Mineev2010,MineevErratum2014,Svirko1994}.

A source of the confusion can be traced to a subtlety in the boundary conditions on the electromagnetic field at the boundary between two media with different gyrotropic constants.  Specifically,  at the interface between two isotropic media,  the requisite boundary conditions are:
\be
\Delta {\bf E}_\parallel ={\bf 0} \ \ {\rm and} \ \ \Delta {\bf B}_\parallel = \alpha\left(\frac { \Delta \gamma}{c}\right)\frac {\partial{\bf E}_\parallel}{\partial t}
\label{bc}
\ee
with $\alpha=1/2$, where $\Delta {\bf E}_\parallel$, $\Delta {\bf B}_\parallel$, and  $\Delta \gamma$ are, respectively, the discontinuity in the parallel components of the electric and magnetic fields, and of the gyrotropy.  The term proportional to $\Delta\gamma$ can be viewed as the contribution of an induced surface current, which for $\alpha=1/2$, but only for this value cancels the contributions of induced bulk currents to the Kerr effect.  We will sketch below an analysis in one simple limit in which the value of $\alpha=1/2$ can be derived from the symmetries of linear response theory.

The linear response in a medium with nonlocal dielectric tensor $\varepsilon_{ij}(\textbf{r},\textbf{r}^\prime;\omega)$  to an electric field $\textbf{E}(\textbf{r};\omega)$ is the electric displacement vector $\textbf{D}(\textbf{r};\omega)$ such that
\be
D_i(\textbf{r};\omega)=\int d^3r^\prime \varepsilon_{ij}(\textbf{r},\textbf{r}^\prime;\omega) E_j(\textbf{r}^\prime;\omega).
\ee
So long as time-reversal symmetry remains unbroken, even in the presence of dissipation, it follows from the fluctuation-dissipation theorem\cite{paulmartinbook} that
\be
\varepsilon_{ij}(\textbf{r},\textbf{r}^\prime;\omega) = \varepsilon_{ji}(\textbf{r}^\prime,\textbf{r};\omega).
\label{dielectric}
\ee

Under the assumption that the dielectric response is short-range ({\it i.e.}, that it falls rapidly as a function of $|\textbf{r}-\textbf{r}^\prime|$), Eq. \ref{dielectric} can be expanded in powers of the range, $a$. To first order in $a$,
\be
D_i= \epsilon_{ij}E_j+\gamma_{ijk}\frac{\partial E_j}{\partial x_k}+\frac{1}{2}\frac{\partial \gamma_{ijk}}{\partial x_k}E_j
\label{gyro}
\ee
where 
\be
\epsilon_{ij}(\textbf{r}) \equiv 
 \int d\textbf{r}^\prime\   \varepsilon_{ij}\left(\textbf{r}+\frac {\textbf{r}^\prime}2 ,\textbf{r} -  \frac {\textbf{r}^\prime}2\right),
\ee
\be
\gamma_{ijk}(\textbf{r}) \equiv 
\int d\textbf{r}^\prime  \  r^\prime_k\ \varepsilon_{ij}\left(\textbf{r}+  \frac {\textbf{r}^\prime}2 ,\textbf{r} -  \frac {\textbf{r}^\prime}2\right).
\label{factor2}
\ee
By defining the moments of $\varepsilon$ relative to the midpoint we have insured that the position-dependent dielectric and gyrotropic tensors preserve the appropriate  symmetries under exchange of indices:
\be
\epsilon_{ij}(\textbf{r}) = \epsilon_{ji}(\textbf{r}); \  \gamma_{ijk}(\textbf{r}) = -\gamma_{jik}(\textbf{r}).
\label{symmetry}
\ee

Note that for Eq. \ref{gyro} to be valid, it is necessary that $a$ be small compared to both the wavelength $\lambda$ and $\ell$, the scale over which the dielectric properties of the medium vary.  In the bulk of a uniform medium, the spatial variation of $\gamma$ can be neglected, leading to the more familiar expression for a gyrotropic medium
\be
D_i(\textbf{r}) = \epsilon_{ij}E_j(\textbf{r})+\gamma_{ijk}\frac{\partial E_j(\textbf{r})}{\partial x_k}
\label{uniformgyro}
\ee
At the interface with another medium, the variation of $\gamma$ cannot be neglected.  However, if the width of the interface is small compared to the wavelength of light, {\it i.e.} $\ell \ll \lambda$, it is possible to treat the medium as uniform, but to apply an appropriate boundary condition at the interface.  While the result  is valid in a much wider range of conditions, in the limit that $a/\ell$ and $\ell/\lambda \ll 1$, where Eq. \ref{gyro} is valid even in the interfacial region, it is straightforward to derive Eq. \ref{bc} by integrating this more general constituitive relation across the interface;  the value of $\alpha=1/2$ arises form the $1/2$ in the last term in Eq. \ref{gyro}

While our attempt to link the observed Kerr effect with naturally active gyrotropy is proven wrong, the phenomenological link between charge ordering (spontaneous breaking of spatial symmetries) and the onset of the Kerr effect is even stronger than it was at the time of publication of Ref. \onlinecite{Hosur2013}. Evidence has accumulated, from NMR\cite{WuYBCONMR} and X-ray\cite{CominLBSCOYBCO, BlackburnXRayOrthoII, LeTacon2014} diffraction studies that density perturbations appear at temperatures coincident with onset of the Kerr rotation. On the other hand, 
while breaking of mirror symmetries 
is certainly necessary to produce a Kerr effect,  
it is not sufficient.  At the time of this writing, we are considering 
 scenarios involving a violation of the one or more of the assumptions of the general argument based on reciprocity, i.e., equilibrium, linear response, and specular reflection. 
 
In particular, regarding the relevance of non-equilibrium effects, it is possible that if such effects were present, they would produce small violations of the symmetries of the response functions, which would otherwise be expected on the basis of the fluctuation-dissipation theorem. Where charge order interacts with quenched randomness, such deviations may not be totally implausible. For instance, transport experiments\cite{CaplanYBCONoise} on YBCO nano-wires have revealed anomalous hysteresis and noise, reminiscent of the non-equilibrium behavior of the random-field Ising model.  Hysteresis in the low field magnetization of various cuprates has also been observed\cite{Panagopoulos2004} with onset temperatures that are comparable to the Kerr onset temperatures.  Whether such non-equilibrium effects would permit a Kerr response even when time-reversal symmetry is present remains an open question. 

Historically, the polar Kerr rotation measurements in the cuprates (along with the observation of quantum oscillations and polarized neutron scattering measurements) were among the first 
experiments  to suggest that there 
is a symmetry-breaking phase transition within the pseudogap regime.  
The subsequent discovery of short-range charge-density-wave order  in NMR and X-ray measurements, both of which appear to have a similar onset temperature as the Kerr rotation, 
provide strong additional support for this notion.  Despite theoretical constraints from reciprocity, it 
appears to us likely that the observation of Kerr rotation and  charge order are linked.  The correct interpretation of the Kerr measurements, which is consistent with the theorems of linear response, and which provides a synthesis of the key phenomenological observations in the pseudogap regime,   remains an important and outstanding theoretical challenge.  

\acknowledgements{ We acknowledge important conversations with P.
Armitage, L. Gorkov, R. Laughlin and V. Mineev.
}

\bibliographystyle{apsrev}
\bibliography{kerr_references}

\end{document}